\documentstyle[preprint,aps]{revtex}

\begin{document}
\draft
\preprint{UM-ChE-98/624}
\title{Site percolation thresholds and universal formulas for the Archimedean lattices}
\author{Paul N. Suding and Robert M. Ziff$^{\dag}$}
\address{Department of Chemical Engineering, University of Michigan, Ann Arbor, MI 48109-2136}  

\date{\today}
\maketitle
\begin{abstract}

The site percolation thresholds $p_c$
are determined to high precision for eight Archimedean (uniform) lattices,
by the hull-walk gradient-percolation simulation technique, with the results
$p_c = 0.697\,043$, honeycomb or ($6^3$),
$0.807\,904$ $(3,12^{2})$,
$0.747\,806$ $(4,6,12)$,
$0.729\,724$ $(4,8^{2})$,
$0.579\,498$ $(3^{4},6)$,
$0.621\,819$ $(3,4,6,4)$, 
$0.550\,213$ $(3^{3},4^{2})$, and 
$0.550\,806$ $(3^{2},4,3,4)$,
and errors of about $\pm 2\times10^{-6}$.
(The
remaining Archimedean lattices are the square $(4^4)$, triangular $(3^6)$ and Kagom\'e $(3,6,3,6)$, 
for which $p_c$
is already known exactly or to a high degree of
accuracy.) \ 
The numerical result for the $(3,12^{2})$ lattice is consistent with the
exact value $[1-2 \sin(\pi/18)]^{1/2}$, which we also derive.
 The  values of $p_c$ for all Archimedean lattices 
are found to be linearly related to the density of 
sites within an error of about 1\%, which is
more accurate  than correlations based solely upon the
coordination number.
Comparison with nonuniform lattices is also made. 
\end{abstract}

\narrowtext
\section{Introduction}
\label{sec1}

The Archimedean (Arch) tilings were first elucidated and published by Kepler (a translation of which can be found in 
\cite{Field}).  These lattices
received their name from references in Kepler's paper to
Archimedes' descriptions of regular solid polyhedra, which are related to these 2d lattices.
The significance of the Arch tilings or lattices,
which are shown in Fig.\ \ref{fig1},
is that they are the complete set of lattices having infinite
tessellation in which all vertices are equivalent.  
(They are also called ``uniform" lattices.)  This
 property has made them useful in the study of mathematics \cite{Fejes}, crystallization  
\cite{Frank,TD}, statistical mechanics \cite{Shrock}, as well as
 percolation \cite{dI}.  The familiar square,
triangular, honeycomb and Kagom\'e lattices are all Arch.

The primary goal of this study was to determine precise values of the site percolation threshold $p_c$ 
for all Arch lattices for which accurate values were not previously known.
A few years ago, d'Iribarne, et al.\  determined $p_c$ for all but one of the Arch
lattices (which they call ``mosaics") to about three significant figures \cite{dI}, using a method 
based upon the analysis of the minimum spanning tree of clusters \cite{Dussert}.
More recently van der Marck (vdM) \cite{Marck97b,Marck98b} determined the thresholds for three of these lattices to
nearly four figures.  Furthermore, the threshold for the honeycomb lattice, a common one for percolation studies,
was previously known only to about four figures \cite{Marck98b,VicsekKertesz,Djor}.
 Here, we extend the precision of these thresholds to six significant figures.
We did not consider the square, triangular, and
Kagom\'e lattices, as $p_c$(site) is either known exactly (triangular and Kagom\'e \cite{ES}),
or has already been measured to a high degree of precision (square \cite{Ziff}) for these cases.

The second goal of this study was to explore the dependence of $p_c$
upon lattice characteristics and determine to what
extent a universal formula can be found for 2d site percolation.  Recently
there has been renewed interest in this question
with the extensive work of Galam and Mauger (GM) \cite{GM4,GM4A,GM6,GM7,GM97c}
and vdM \cite{Marck97b,Marck98b,Marck97a,Marck97c,Marck98a}.  The Arch lattices provide an
excellent resource for studying
this question because they possess a range of coordination numbers from 3 to 6 and
a variety of other lattice characteristics.

\subsection{Nomenclature}
Unfortunately, neither Kepler nor Archimedes (nor anyone else to our knowledge) 
assigned common names to all the Arch
lattices.  However, Gr\"unbaum and Shephard \cite{GS}
have introduced a general notation, which we use here, to catagorize such lattices
in terms of
the  set of polygons which surround each vertex, $(n_1^{a_1}, n_2^{a_2}, \ldots)$.
Going clockwise around
a vertex, the numbers $n_i$ denote the number of sides of each polygon,
and the superscript $a_i$ refers to the number of these
polygons adjacent to each other.   
For example, the triangular lattice has six triangles around a given point, and is designated
$(3^6)$.
The labels for all the Arch lattices are shown in Fig.~1.
Note that the $(3,12^2)$ lattice has been 
called the ``3-12" lattice in the literature, and in places the $(4,8^2)$ has been given the
name of ``bathroom tile" lattice.

For convenience in our work we did assign nicknames to the lattices,
which are listed in Table \ref{table1}.  The
$(4,6,12)$ is called the ``cross" lattice because of the
cross in  Fig.\ \ref{fig3}; the $(4,8^2)$ lattice is named after our local
shopping mall, Briarwood, which (like many such malls) has tiles of this shape on its floor;
$(3^3,4^2)$ is called the ``direct" lattice because of its
directed  structure, etc.    The origins of the rest of the names are, we hope, somewhat more obvious.
\section{Method}
\label{sec2}

\subsection{Hull Gradient Walk}

To find $p_c$ we employ the hull-gradient method \cite{ZiffSapoval}, which we previously used to determine 
$p_c$ for site percolation on the square lattice \cite{ZiffStell} and for bond percolation on the Kagom\'e lattice
\cite{ZiffSuding}, both to more than six significant digits of precision.  

In gradient percolation \cite{SRG,RGS}, a linear
gradient of occupied sites is applied in the vertical direction of the lattice.
As the height increases, the concentration of
occupied sites also increases.  This gradient forms two ``land masses'' within the lattice: a continuous region of
occupied sites located at the top of the lattice and a continuous region of vacant sites located at the bottom (Fig.\ 
\ref{fig2}).  The boundary between these two regions forms a path whose average height provides an estimate for
$p_c$.

To make this method efficient for finding $p_c$, we employ a hull-generating
walk \cite{ZCS,WT} which simultaneously generates and identifies the interfacial boundary.
The status  (whether occupied or
vacant) of a site encountered by the walk is determined by generating a random number and comparing that number with the
occupation probability for that height.  If the walk does not arrive at a given site,
the status of that site will not
be determined, and no random number will be generated for it, contributing to the efficiency.
In fact, we believe that the hull-gradient method is the most
efficient way to determine $p_c$ for 2d lattices.
It is also quite simple to program, as it involves
no lists or cluster labelling algorithms.

For each gradient $|\nabla p|$, an estimate of $p_c$ may also be obtained
by taking the ratio of occupied to total
 (occupied plus vacant perimeter) sites belonging to the hull \cite{RGS}.  As the size of the system is
increased or the gradient is decreased, $p_c(|\nabla p|)$ approaches $p_c$ linearly in $|\nabla p|$.  
Thus a simple extrapolation of the data for finite gradients gives the infinite-system
value.  In practice, we considered
gradients sufficiently small (typically of the order of $10^{-4}$ $\Delta p$/lattice spacing) that the extrapolation
from the final point to zero gradient was of the order of the error bars of that last point.
Additional details of the method are given in \cite{ZiffSapoval,ZiffSuding}.

\subsection{Computer Techniques}

In the simulation, the status of all points visited by the walk must be stored in computer memory.  To accomplish 
this efficiently, all of the Arch lattices were transformed to align on
rectilinear grids, as shown in Fig.\ \ref{fig3}.
Note that the gradient was not applied directly to these squared-off forms,
as some lattices had to be distorted in the vertical direction to get them in 
this form.
Instead, all numerical calculations were performed so that the height of the point in the original, undistorted, lattice was
used to find $p(z)$.  Previously we
showed, for bond percolation on the Kagom\'e lattice,
that assigning $p(z)$ this way yields the best linear scaling of $p_c$ vs.\
gradient \cite{ZiffSuding}.  For two lattices, we considered two possible orientations of the
lattice and found similar results, as discussed in Appendix A.

The lattices were initialized by filling the top half of the first column with occupied sites and the lower half with 
vacant sites, which prevents the walk from closing on itself at the start of the algorithm (Fig.\ \ref{fig2}).  Periodic
boundary conditions were applied in the horizontal direction, and each new column to the right was cleared as it was
visited.  This allowed the simulation to run indefinitely and have essentially no boundary effects from the horizontal ends
of the system.  The maximum distance the walk traveled horizontally from the front was tracked to detect if wraparound
errors occurred; if they did, the system size and/or gradient was adjusted accordingly, and the run was restarted with
the expanded system.

Random numbers were generated using the shift-register sequence generator R7(9689) \cite{Ziff98,Gol} defined by:
\begin{equation}
x_n = x_{n-471} \ ^\wedge \ x_{n-1586} \ ^\wedge \ x_{n-6988} \ ^\wedge \ x_{n-9689} \ ,
\label{rng}
\end{equation}
where $^\wedge$ is the bitwise exclusive-or operation.  This `four tap' generator is equivalent to decimating by 7
(taking every seventh term)  of the sequence generated by the two-tap rule R(9689)
 ($x_n = x_{n-471} \ ^\wedge \ x_{n-9689}$) given by
Zierler \cite{Zier}, where the
decimation has the effect of greatly reducing the three- and four-point correlations of the two-tap generator \cite{Ziff98}.
Our previous work has shown that
this generator does not appear to
introduce  errors in simulations of this kind \cite{ZiffSapoval,ZiffSuding}.

\section{Results and Discussion}
\label{sec3}

\subsection{Percolation thresholds}

For each lattice, $p_c$ was plotted as a function of the gradient.  Fig.\ \ref{fig4} shows a representative plot for the
$(3,12^{2})$ lattice.
In all cases, a linear relationship was observed between
the magnitude of the gradient and the estimate $p_c(|\nabla p|)$, as found previously
for the square \cite{RGS} and Kagom\'e lattices \cite{ZiffSuding}.  The value 
at the intercept of the $y$-axis represents $p_c$ for an infinite lattice (zero gradient), and these values are reported in Table
\ref{table1}.  Also shown in Table \ref{table1} is the quantity of random numbers generated
for each simulation, which is identical to the total number of occupied and vacant sites belonging to
the hull, because one random number is generated for each new site visited.   These simulations
consumed several months of computer time on Sun and HP workstations.

For all our values, the error (which is purely statistical
in origin) is of order $10^{-6}$.  Our results are
consistent with d'Iribarne et al.'s three-digit values, if we 
assume an error of $\pm 0.003$ in their data.
Our result $0.697\,043(2)$ for the honeycomb lattice is
consistent with the previous values 
0.6973(8) of Vicsek and Kert\'esz \cite{VicsekKertesz},
0.6962(6) of Djordjevic et al.\ \cite{Djor}, and
0.6971(2) of vdM \cite{Marck98b}.
Likewise, our results agree with the determinations 0.5504(2) for $(3^3,4^2)$ 
and 0.7298(1) for $(4,8^2)$ made recently by vdM \cite{Marck97b,Marck98c}.

In another recent work, Pr\'ea \cite{Prea} has calculated the distance sequences $c_n$, 
defined as the number of sites $n$ steps (chemical distance) 
from the origin,  for the 
Arch lattices.  Pr\'ea found a monotonic relation between
$p_c$(site) and $c \equiv \lim_{n \to \infty} \inf (c_n/n)$, although three different
lattices with different $p_c$ --- the square, Kagom\'e, and $(3,4,6,4)$ ---
share the same value $c=4$. 
This apparent monotonic behavior led Pr\'ea to conjecture that
$p_c$ for the $(3^4,6)$ lattice (the one Arch lattice not studied previously)
is within in interval $(0.55,0.6)$.
Our result 
$0.579\,498(2)$ confirms this conjecture.

\subsection{Exact percolation threshold for the ($3,12^{2}$) lattice}

The threshold for the  $(3,12^{2})$ can be derived exactly.  The proof is based upon the similarity of
structure between this lattice and the Kagom\'e lattice.  In Fig.\ \ref{fig5}, the $(3,12^{2})$  lattice  is shown with
the triangle bonds in bold and the bonds between the triangles as dashed lines. If the dashed lines are reduced
in length, the $(3,12^{2})$  lattice transforms into the Kagom\'e lattice.  Similarly, the sites on both ends of the dashed
line in the
$(3,12^{2})$  (marked by the large circle in the figure), taken together, are equivalent to a single point on the
Kagom\'e lattice.

Let both lattices be at the percolation threshold and $p$ be the probability that a given site on the $(3,12^{2})$  lattice
is occupied.  Then, $p^{2}$ is the probability that both sites at the ends of the dashed bonds
 are occupied, and the bond will permit flow. The
dashed bond is equivalent to a single point on the Kagom\'e.  At the percolation threshold, the probability that a site is occupied
on the Kagom\'e lattice is $1 -2 \sin (\pi/18)$.  Therefore, for the ($3,12^{2}$) lattice,
\begin{equation}
p_c = [1-2\sin(\pi/18)]^{1/2} = 0.807\,900\,764\ldots \ .
\label{exact}
\end{equation}
Our numerical estimate $0.807\,904$  is consistent with this result within the error bars
$\pm 0.000\,003$.  This agreement provides  a further validation of both our algorithm and error analysis.
 
\subsection{Galam and Mauger's universal formula for $p_c$}

Early in the work of percolation, thresholds were related with coordination number $q$ using relatively simple 
formulas.  More recently, GM \cite{GM97c,GM7}  have attempted to refine this by introducing
different classes
of lattices, characterized by the constants $p_0$, $a$ and $b$ in their formula
\begin{equation}
p_c = p_0[(d-1)(q-1)]^{-a}d^{b} \ ,
\label{Galam}
\end{equation}
where $d$ is the dimensionality. 
By defining two different classes, each with distinct sets of constants $p_0$ and $a$ for site
and bond percolation  (with $b=a$ for bond and $b=0$ for site percolation), GM were
able to obtain a fairly good fit of $p_c$ for a number of lattices of
various dimensionality, with a fifth set of constants $p_0$ and $a$
required for systems of very high dimensionality.
Their class 1 contains all 2d lattices except the Kagom\'e, while class 2 comprises 
all higher-dimensional systems plus the 2d Kagom\'e.

Subsequently, vdM \cite{Marck97b,Marck97a} considered a number of additional lattices and argued 
that a formula of this type, depending only upon $d$ and $q$,
cannot be very accurate because of the existence of different lattices with differing
$p_c$ but identical $d$ and $q$.
GM \cite{GM7} responded that
their formula is applicable only for ``isotropic" lattices in which each site has
the identical $q$.  (In this present paper, we describe these lattices as ``uniform,"
by which we mean that each vertex is identical in terms of its surrounding polygons, allowing
for rotations and reflections.)  
For other lattices, GM  
reformulated (\ref{Galam}) by reinterpreting $q$ to be an effective coordination number $q_{\rm eff}$
(not necessarily the average value $\overline q$).  Because $q_{\rm eff}$ cannot  be found independently,
this reinterpretation has the effect
of turning their formula into a correlation between site and
bond thresholds for a given lattice rather than a more general correlation for $p_c$, which
is developed further in \cite{GM98} and discussed in \cite{Marck98b,Babalievski}.  However, because we do not consider 
thresholds for bond percolation here, we do not address this aspect of GM's work in this paper.

In Fig.\ \ref{fig6} we plot our results using the same axes as GM
\cite{GM97c}, ln($1/p_c$) vs.\ ln($q-1$).  We also exhibit GM's
two fitting formulas for site percolation, which are straight lines on 
this plot.
Clearly, many of the Arch lattices do not fall near either of these two
lines.  Evidently, GM's approach (\ref{Galam}) cannot accommodate these additional uniform lattices,
unless one introduces still more lattice classes, which is
counter to the concept of universality.

\subsection{A new fitting formula for $p_c$ \label{section:formulas}} 

For the Arch lattices, one can observe the trend that $p_c$ increases with increasing ``openness" of the
lattices.  To characterize the latter, we introduce the quantity
$\rho$ equal to the density or number of lattice
sites per unit area, assuming that all 
bond lengths are unity.  Its inverse,
the area per site, can be determined by drawing lines which bisect
the centers of each polygon surrounding a given site, as in a dual-lattice
construction, and summing the enclosed area.  The tiles defined by these bisector lines clearly  
fill the entire lattice.
The total area per site is simply
 1/3 the area of each triangle, 1/4 the area of each square, etc.,
surrounding that site.  
For Arch lattices with uniform bond lengths, all polygons are regular.
The area of 1/$n$-th of a 
regular $n$-gon, with unit edge length, is given by
\begin{equation}
A_n =  {1 \over 4} \cot {\pi \over n} \ , 
\label{area}
\end{equation}
with some numerical values
given in Table \ref{table4}.
For a lattice characterized by vertices $(n_1^{a_1}, n_2^{a_2},\ldots)$,
we then have
\begin{equation}
\rho = \left[ \sum_{i} {a_i} A_{n_i}\right]^{-1}
= 4 \left[ \sum_{i} a_i \cot {\pi \over n_i} \right]^{-1}  \ .
\label{areatotal}
\end{equation}
The resulting values of $\rho$ for the Arch lattices 
are listed in Table \ref{table2}. 
In  Fig.\ \ref{fig7} we show a plot of $p_c$ vs.\ $\rho$, and indeed
one can see that the correlation between $p_c$ and $\rho$ is good.  Furthermore,
that correlation can be well
represented by a simple linear function.
A least-squares fit yields
\begin{equation}
p_c =  1.0405 - 0.4573 \rho \ . \label{fit}
\end{equation}
The errors involved in using this formula are
listed in the column $\Delta$ in Table \ref{table2},
and are generally  within about 0.01,
in contrast to errors up to 0.05
if GM's correlations were used (taking the best class in each case).
In Fig.\ \ref{fig8} we plot the errors from using these different fitting formulas
for the various lattices, clearly showing how GM's formulas are each accurate for
certain lattices only, while (\ref{fit}) works fairly well for all of the lattices.
The rms errors from using (\ref{fit}) for the 11 Arch lattices equals 0.0075.

It turns out that our fit (\ref{areatotal}-\ref{fit}) is related to
a correlation for $p_c$ given by Scher and Zallen nearly three decades ago  
\cite{ScherZallen}.  These authors introduced a ``filling factor" $f$ defined as the fraction
of space occupied by disks of radius $1/2$  placed at each lattice site, 
and found for four lattices in 2d,
\begin{equation}
fp_c \approx {\rm const.}  = 0.44 \pm 0.02 \ .  \label{SZ}
\end{equation}
Now, $f$ is simply $\pi \rho /4$, so
that their correlation also relates to the density
of sites.
However, while their hyperbolic relation (\ref{SZ}) provided a
good fit to lattices they considered (to the accuracy available
at that time), it does not capture the behavior for the other lattices, especially those
of low $\rho$, as shown in Fig.\ \ref{fig7}, where (\ref{SZ}) is also plotted.

Fairly recently, d'Iribarne et al.\ \cite{dI} considered a  correlation for $p_c$ of the 
Arch lattices that is also closely related to $\rho$ or $f$.
They compared $p_c$ to a variable $m$
which represents the average edge length of a minimally spanning tree on a complete
lattice, normalized by the area.  Because on the complete graph there is one edge
per site, $m$ is identical to $\rho^{1/2}$.
Their plot of $p_c$ vs. $m$ correlates the data to a single curve just
as our Fig.\ \ref{fig7} does, however, in terms of $m$ the behavior is quite curved,
which d'Iribarne et al.\  fit to the quadratic
\begin{equation}
p_c = 0.685 + 0.799 m - 0.899 m^2 \ .  \label{dIRR}
\end{equation}
In fact, this fit reduces the rms error of our results over the 11 lattices to 0.0057. 
(Note, only a slight improvement 
is obtained by adjusting these coefficients for our new values of $p_c$.)  However,
the expense of this improvement is  an
extra parameter in (\ref{dIRR}) over (\ref{fit}).
Indeed, one can also extend (\ref{fit}) to a quadratic in $\rho$,
with the result
\begin{equation}
p_c  =  0.9472 - 0.2181 \rho - 0.1439 \rho^2 \label{quadfit} \ ,
\end{equation}
which gives a comparable rms error over the 11 lattices, 
0.0054, but again at the expense of a more complicated formula than (\ref{fit}).

Along these lines, it is interesting to note that an excellent fit can be achieved
using results from only the three lattices where $p_c$ is known exactly (which
happen to span the whole range of $p_c$ values).    That fit,
\begin{equation}
p_c  =  0.9466 - 0.1972 \rho - 0.1642 \rho^2 \ , \label{exactfit}
\end{equation}
 yields an rms error over the 11 Arch lattices of 0.0062,
only slightly worse than (\ref{dIRR}) or (\ref{quadfit}).
This fitting formula depends upon no adjustable
parameters nor any Monte-Carlo-measured values.

\subsection{Extension to non-uniform lattices}
We have found that (\ref{areatotal}-\ref{fit}) may also be applied to 
non-uniform lattices, if in calculating $\rho$ by (\ref{areatotal}),  $a_n$ is
interpreted as the {\it average} number of polygons of type $n$
over all vertices. 
For the non-uniform lattices (also referred to as anisotropic \cite{GM97c}), many of the 
polygons are not regular, and it is not always possible to make all
bonds of the same length, which were the two assumptions in deriving 
(\ref{areatotal}).  Thus, we now consider (\ref{areatotal}) to represent
a topological or graphical weight rather than an actual geometric property of the lattices.

We looked at seven non-uniform lattices considered by
vdM \cite{Marck97b,Marck98b} and/or  GM \cite{GM97c} and
whose values of $p_c$ are known (we did not determine any of these $p_c$'s here):
the bowtie, bowtie dual,
pentagonal, $(4,8^2)$-matching, dice, 
Penrose, and Penrose dual lattices.
The first five of these lattices are shown in Fig.\ \ref{fig:nonuniform}, while the
quasicrystalline Penrose and Penrose dual are shown for example in \cite{YonezawaSakamotoHori}
and \cite{GM97c}.  The form of the Penrose lattice used here is the rhomb or P3 form \cite{GS}.

For the dice lattice, ${1\over3}$ of the vertices are $(4^6)$ while
${2\over3}$ of them are $(4^3)$, so on the average the vertices are
$(4^4)$.  Likewise for the $(4,8^2)$-matching, we have
${1\over3}(3^2,8^2)\ +$ ${2\over3}(3^2,4,8) =$ $(3^2,4^{2/3},8^{4/3})$, for the bowtie 
${1\over2}(3^4,4^2)\ +$ ${1\over2}(3^2,4^2) =$ $(3^3,4^2)$, 
for the bowtie dual  ${1\over3}(4^2,6^2) \ +$ ${2\over3}(4,6^2) =$ $(4^{4/3},6^2)$, and
for the pentagonal  ${1\over3}(5^4) \ +$ ${2\over3}(5^3) =$ $(5^{10/3})$.
For the Penrose lattice,
the average vertices are simply $(4^4)$, while for the 
Penrose dual lattice the average vertices are \cite{LuBirman}
$(3^{a_3},4^{a_4},5^{a_5},6^{a_6},7^{a_7})$ 
with $a_3 = 3  (x^2 + x^4) = 3  (3-4x) = 1.583592135$,
$a_4 = 4  x^5 = 4(5x-3)  = 0.360679775$,
$a_5 = 5  (x^3 + x^6) = 5(4 - 6x) =1.458980338$,
$a_6 = 6  x^7 = 6(13x-8)= 0.206651122$, and
$a_7 = 7  x^6 = 7 (5-8x)= 0.39009663$,
where $x = (\sqrt{5}-1)/2$ is the inverse of the golden ratio.
Note that in general $a_n$ satisfy $\sum_n (1-1/n) a_n = 1$  and  $\sum_n a_n = \overline q$,
where $\overline q$ is the average coordination number.

These average vertex numbers are listed in 
Table \ref{table3}, along with 
 $p_c$ values,
$\rho$ values, $p_c^{\rm est.}$ and error $\Delta$ from using  
 (\ref{areatotal}-\ref{fit}).
Interestingly, the $\Delta$ for all 7 of
these lattices are generally smaller that those for
the Arch lattices --- with an rms value of 0.0046
vs.\ 0.0075.
(Note that some of these $\Delta$'s are however close to the precision of
the value of $p_c$.)  Surprisingly, even the Penrose dual lattice, which incorporates polygons
with from 3 to 7 sides and a very irregular structure, is also well modelled by this formula.
Evidently, this approach, which is based upon $\rho$ given by  
(\ref{areatotal}), appears to be quite robust.

One could conceivably use these additional data points from the non-uniform 
lattices to refine the
fit of the  formulas for $p_c(\rho)$ in Section \ref{section:formulas}.
However, the improvement turns out to be marginal, and considering the
relatively low precision of
the values of $p_c$ for the non-uniform lattices, the justification for doing this is doubtful.

Note that if
one retained the definition of $\rho$ literally as the density of sites
for these lattices, rather than using the definition above, 
then the fit of the non-uniform lattices with (\ref{fit}) would be very poor.  An example is given by
the dice lattice, which has a site density identical to that of the 
triangular lattice, $2 \sqrt{3}/3 = 1.1547$.  The $p_c$ for this lattice, 0.5851 \cite{Marck97b},
is substantially above the value for the triangular, 1/2, and the data point for this
case would fall well above the line in Fig.\ \ref{fig7}.

It is also interesting to note that we have now considered
three different lattices with the same average
vertex environment of three triangles and two squares: the two Arch lattices 
$(3^3,4^2)$ and $(3^2,4,3,4)$, and the non-uniform bowtie lattice,
which have $p_c$'s of $0.550\,806$, $0.550\,213$, and $0.5475(8)$ respectively.
The closeness of these values supports the idea that
$p_c$ is determined principally by the local nature of the lattice as 
characterized by the vertex numbers.  
For the two Arch lattices, the $p_c$'s are nearly identical, and close
to the prediction by (\ref{fit}) of $0.5503$.
The bowtie lattice, which has the same vertex numbers as the others
on the average only,  has a value of $p_c$ that is slightly lower,
by about 0.003.

The inadequacy of using $q$ alone to estimate $p_c$ is also apparent, as
the $(3^4,6)$ lattices has the same $q=5$ as these three above, but a $p_c$
that is substantially higher, $0.579\,498$.
The results for the non-uniform lattices are also included in Fig.\ \ref{fig6},
using $\overline q$ for $q$.

A similar situation occurs for the square ($p_c = 0.592\,746$), dice ($p_c = 0.5848$), and 
Penrose ($p_c = 0.5837$)
lattices, whose (average) vertices are $(4^4)$ in all cases.  
Again, the non-uniform lattices have
a slightly lower $p_c$ than the uniform one with the same vertices,
with the ``most" non-uniform one (the Penrose) having the lowest $p_c$.
And again, other lattices with $q=4$ but different average vertex numbers (there are four of them here)
all have much
different (and higher) $p_c$'s. 

van der Marck has pointed out \cite{Marck97c,Marck98c} that adding an extra site at the center
of any triangle on a 2d lattice does not change the value of $p_c$ (site).  If these extra
sites are included in the calculation of $\rho$, then $\rho$ would change and
the predictions of \ref{fit} would be poor.
Therefore, in applying these formulas, one must disregard these superfluous sites.
However, there are undoubtedly other classes of lattices where \ref{areatotal}-\ref{fit} 
will not work well --- and will not be so easily fixed.

\subsection{Further variations on the fitting formula}

Although a precise universal formula for $p_c$ based upon the vertex numbers alone
is impossible, it is interesting to 
explore  variations to (\ref{areatotal}-\ref{fit}) to see what improvements can be made.  According to
(\ref{areatotal}), the variable $\rho^{-1}$ is an average over
the nearest neighbors polygons weighted by $A_n$.
To see how close this 
weighting compares to an optimal one, we carried out a regression analysis,
allowing all $A_n$ as well as the linear fit of $p_c(\rho)$ to vary,
and minimizing the error from the linear fit using data from all 18 lattices given above, both Arch
and non-uniform.
This procedure yielded
the $A'_n$ given in Table \ref{table4} 
and the fit
\begin{equation}
p_c =  1.1383 - 0.5503 \rho' \ ,  \label{fit2}
\end{equation}
where $\rho' = \sum a_n A'_n$.
Note that we kept $A'_4 =
0.25$, as one weight can be fixed arbitrarily.

The errors using this fit are generally lower than those that result from (\ref{fit});
in fact the rms error over all 18 lattices decreases substantially from 0.0065 when (\ref{fit}) is used
to 0.0028 when (\ref{fit2}) is used.  
However, this formula is
clearly less predictive than eq.\ (\ref{fit}),
which had no adjustable parameters in the definition of $\rho$.

The $A'_n$ listed in Table \ref{table4}
fall close to, but in a somewhat narrower range than, the $A_n$. 
(Note $A'_7$ is somewhat out of line with the rest, being bigger than $A_7$, but
was derived from only one lattice, the Penrose dual.) \   
These results suggest that there might
be a topological weighting rule that is better than the effective area per site, (\ref{areatotal}). 
However, we do not know what that weighting rule may be.

To test if a non-linear relation between $p_c$ and $\rho'$ may be better, we carried out
the same regression analysis as above, but allowed for a quadratic $p_c(\rho)$.
Interestingly, the $A'_n$ came out nearly the same, and the behavior of $p_c(\rho)$
was very nearly linear.

Using d'Iribarne et al.'s form (\ref{dIRR}), with $m = \sqrt{p'}$ as the variable and assuming
a quadratic relation between $p_c$ and $m$, we found optimal weights $A'_n$ now much closer to
the $A_n$.  These various and partly contradictory results suggest that, while some improvements
can be made, our original
assumptions --- that $A_n$ is the actual polygon area, and that $p_c(\rho)$ is linear --- are not 
 unreasonable.

Of course, there
are deeper assumptions to this whole approach --- that $\rho^{-1}$ is a linear function  
of the $a_n$, and that $p_c$ is a simple linear or quadratic function of $\rho$ ---
which we  have not explored.
And ultimately, to get a more precise formula for $p_c$, one
must also take into account the
actual arrangement of the polygons around each vertex, not just the number of each type of polygon.
Clearly, further work can be done in understanding the relation of $p_c$ to lattice type.

\subsection{Conclusions}

We have determined the $p_c$(site) for the honeycomb and seven less-common
Arch lattices to nearly six significant figures,
about 100 to 1000 times more precise than previously known --- except for
the $(3^4,6)$ lattice, whose $p_c$ was never determined previously.  The value for the latter lattice is within
the conjectured bounds of Pr\'ea \cite{Prea}.  We find a fairly good linear correlation between $p_c$
and $\rho$ (and related to the correlations of Scher and Zallen \cite{ScherZallen} and of d'Iribarne et al.\ \cite{dI}), although 
its accuracy is far below what can be found in simulation.  This correlation can be extended
to non-uniform lattices, and can be improved somewhat if the weightings $A_n$ are adjusted to the 
$A'_n$.  The correlation with $\rho$ is substantially more accurate than correlations
based primarily upon coordination number, such as those of GM \cite{GM4,GM4A,GM6}.   

All our results here are restricted to 2d and to site percolation.  It would be interesting to see
if these ideas on relating $p_c$ to the local graphical environment can be extended to bond 
percolation and to higher dimensionality.  It would also be interesting to see how well GM's ideas relating
$p_c$(site) and $p_c$(bond) on the same lattices hold for the Arch lattices.

\acknowledgments

The authors thank Steven van der Marck for comments and for providing
unpublished improved $p_c$ values for some non-uniform lattices.
This material is based upon work supported by the US 
National Science Foundation under Grant No.\thinspace DMR-9520700, and by the Shell Oil Company Foundation.

\appendix
\section{Importance of gradient orientation}

In order to verify that the direction that the gradient is applied on a lattice has no 
effect on $p_c$, we calculated $p_c$ for both the honeycomb and $(3^{3},4^{2})$
lattices after they had been rotated 90 degrees.  Fig.\ \ref{fig9} shows the original orientation  of the simulation and
the orientation after the lattice has been rotated 90 degrees.  The $p_c$ for the honeycomb lattice was measured as
$0.697\,043 \pm 0.000\,002$ and then $0.697\,046 \pm 0.000\,003$ after 
90 degree rotation.  Likewise, for the $(3^{3},4^{2})$
lattice, $p_c$ was measured as $0.550\,213 \pm 0.000\,002$ and then $0.550\,211 \pm 0.000\,002$ after 90 degree rotation. 
In both cases, the $p_c$ values were identical within the error of the method, verifying that the orientation of the
lattice has no effect on determining $p_c$ in our method.

$^{\dag}$Electronic mail: rziff@engin.umich.edu

\begin{figure}
\caption{The 11 Archimedean (Arch) lattices, in which all vertices are equivalent. Lattices are designated
using the notation of Gr\"unbaum and Shephard [23] as explained in the text.
\label{fig1}} 
\end{figure}

\begin{figure}
\caption{ 
The hull-generating walk along a percolation gradient for the honeycomb lattice.
 Filled circles denote occupied sites and
heavy lines show the hull of the percolating region.  The arrow
points to the starting point of the walk.
\label{fig2}}
\end{figure}

\begin{figure}
\caption{ 
The Arch lattices transformed to a square array, for use in the computer simulation.
Some lattices were
distorted in the vertical direction, but in all cases the actual lattice height, rather than the height in the square array, was used to determine
the site occupation probability. 
\label{fig3}}
\end{figure}

\begin{figure}
\caption{Percolation threshold vs.\ lattice gradient for the $(3,12^{2})$ lattice.  As the gradient becomes smaller, the
lattice approaches infinite size.  The percolation threshold can be estimated from the $y$-intercept of a linear regression
of results from finite lattices.
\label{fig4}}
\end{figure}

\begin{figure}
\caption{Derivation of the exact percolation threshold for the $(3,12^{2})$ lattice.  Replacing the dashed bonds with single points leads to 
the Kagom\'e lattice.
\label{fig5}}
\end{figure}

\begin{figure}
\caption{Thresholds plotted as ln $1/p_c$ vs ln $(q-1)$.
($\bullet$) regular lattices (from left to right: honeycomb, square
and triangular),  (\rlap{$\sqcup$}$\sqcap$) Kagom\'e,
($\triangle$) less-common Archimedean
lattices, ($\times$) non-uniform lattices.
Dashed line: GM's class 1
formula; solid line, GM's class 2 formula.  Evidently, many 
lattices do not conform to GM's classification scheme.
\label{fig6}}
\end{figure}

\begin{figure}
\caption{Percolation thresholds as a function of $\rho$,
with the same symbols as in Fig.\ \ref{fig6}.
The linear fit
is given in eq.\  (\ref{fit}).  Also shown are is correlation
of Scher and Zallen, eq.\ (\ref{SZ}) (dashed line).
\label{fig7}}
\end{figure}

\begin{figure}
\caption{Error from using GM's class 1 formula (\rlap{$\sqcup$}$\sqcap$),
class 2 formula  ($\triangle$), and our fit
(\ref{areatotal}-\ref{fit}) ($\bullet$), for the 18 lattices, plotted sequentially in the order they
are listed in Tables
\ref{table2} and \ref{table3} respectively.
Formulas (\ref{dIRR}), (\ref{quadfit}) or (\ref{fit2}) improves the fit 
of (\ref{areatotal}-\ref{fit}) even further.
\label{fig8}}
\end{figure}

\begin{figure}
\caption{Five of the non-uniform lattices used to test the fitting formulas.
\label{fig:nonuniform}}
\end{figure}

\begin{figure}
\caption{Two orientations for the $(3^{3},4^{2})$ and honeycomb lattices
used to test the influence of lattice orientation in our method
of determining $p_c$.
\label{fig9}}
\end{figure}

\vfill\eject

\begin{table}
\caption{Measured values of $p_c$ for site percolation on the Archimedian lattices.
Lattice nicknames in quotes are ours.
\label{table1}}
\begin{tabular}{llc}
Lattice&Measured $p_c$&Total Random Numbers Generated\\
\tableline 
\tableline
$(3,12^{2})$ ``star" &$0.807\,904(3)$&$4.1 \times 10^{11}$\\
$(4,6,12)$ ``cross" &$0.747\,806(3)$&$2.6 \times 10^{11}$\\
$(4,8^{2})$ ``Briarwood" &$0.729\,724(2)$&$2.6 \times 10^{11}$\\
($6^{3}$) honeycomb &$0.697\,043(2)$&$4.0 \times 10^{11}$\\
$(3,4,6,4)$ ``bounce" &$0.621\,819(2)$&$2.9 \times 10^{11}$\\
$(3^{4},6)$ ``bridge" &$0.579\,498(2)$&$2.9 \times 10^{11}$\\
$(3^{2},4,3,4)$ ``puzzle" &$0.550\,806(2)$&$2.8 \times 10^{11}$\\
$(3^{3},4^{2})$ ``direct" &$0.550\,213(2)$&$2.9 \times 10^{11}$\\
\end{tabular}
\end{table}

\begin{table}
\caption{Weights $A_n$  (\ref{area}),  and optimized weights
$A'_n$ which yield the best linear fit of $p_c(\rho')$ over all 18 lattices.
\label{table4}} 
\begin{tabular}{ccc}
$n$ & $A_n$ & $A'_n$\\ 
\tableline
\tableline
3 & 0.144338 & 	0.14400\\
4 &  0.25	 & 	0.25	\\
5 &  0.344095 &  0.33640\\
6 &  0.433013  & 0.41653\\
7 &  0.519130  & 0.52539\\
8 & 0.603553	 & 0.55041\\
12 & 0.933013  & 0.75490\\
\end{tabular}
\end{table}

\begin{table}
\caption{Values of $p_c$ and lattice characteristics for all 11 Arch lattices.
$\Delta$ is the error when (\ref{fit}) is used to fit the data.
\label{table2}}
\begin{tabular}{lcccccc}
Lattice&$p_c$  &Ref.&$ q $ & $\rho$ by (\ref{areatotal}) &$p_c^{\rm est.}$  by  (\ref{fit}) & $\Delta$  \\
\tableline
\tableline
$(3,12^{2})         $&$  0.807\,904    $&&            3  &  0.4974   &   0.8130    &$   -0.005 $ \\
$(4,6,12)           $&$  0.747\,806    $&&            3  &  0.6188   &   0.7575    &$   -0.010 $ \\
$(4,8^{2})          $&$  0.729\,724    $&&            3  &  0.6863   &   0.7266    &$    0.003 $ \\
$(6^{3})  $ honeycomb&$  0.697\,043    $&&            3  &  0.7698   &   0.6885    &$    0.009 $ \\
$(3,6,3,6) $ Kagom\'e&$  0.652\,703\,6 $ &\cite{ES} & 4  &  0.8660   &   0.6444    &$    0.008 $ \\
$(3,4,6,4)          $&$  0.621\,819    $&&            4  &  0.9282   &   0.6160    &$    0.006 $ \\ 
$(4^{4})$ square &$  0.592\,746\,0 $ &\cite{Ziff} &   4  &  1        &   0.5832    &$    0.010 $ \\
$(3^{4},6)          $&$  0.579\,498    $&&            5  &  0.9897   &   0.5879    &$   -0.008 $ \\
$(3^{2},4,3,4)      $&$  0.550\,806    $&&            5  &  1.0718   &   0.5503    &$    0.0005  $ \\
$(3^{3},4^{2})      $&$  0.550\,213    $&&            5  &  1.0718   &   0.5503    &$   -0.0001 $  \\
$(3^{6}) $ triangular &$  0.5           $ &\cite{ES} & 6  &  1.1546   &   0.5124    &$   -0.012 $ \\
\end{tabular}
\end{table}

\begin{table}
\caption{Data for non-uniform lattices.  Here vertex numbers represent averages.
\label{table3}}
\begin{tabular}{lcccccc}
Lattice&$p_c$   &Ref. & $ \overline q $ &  $\rho$ by (\ref{areatotal}) &$p_c^{\rm est.}$  by  (\ref{fit}) & $\Delta$ \\ 
\tableline
\tableline
$(4,8^2)$--matching  $(3^2,4^{2\over3},8^{4\over3})$ &$0.6768(2)$&\cite{Marck97b,Marck98c} & 4 & 0.7936 & 0.6776 & $
-0.0008
$ \\ bowtie dual $(4^{4\over3},6^2)$  &$0.6649(2)$ &\cite{Marck97b,Marck98c} & $3{1\over 3}$ & 0.8338   &   0.6592 
&   $ 0.006   $
\\ pentagonal $ (5^{10\over3})$ &$0.6476(2)$&\cite{Marck98b,Marck98c} & $3{1\over 3}$  &  0.8719  & 0.6418 & $ 0.006  
$\\ Penrose dual (see text) &$0.6381(3)$ &\cite{YonezawaSakamotoHori} & 4  &  
0.8987   &   0.6295   &   $ 0.009   $ \\
dice $(4^4)$&$0.5848(2)$
&\cite{Marck97b} & 4  &   1     &   0.5832   &   $  0.002   $ \\
Penrose
$(4^4)$&$0.5837(3)$ &\cite{YonezawaSakamotoHori} & 4  &   1    &   0.5832   &  
$  0.0005$\\ bowtie $\bowtie$ $(3^3,4^2)$&$0.5474(2)$&\cite{Marck97b,Marck98c} &   5  &  
1.0718   &  0.5503  &  $ -0.003 $ \\
\end{tabular}
\end{table}
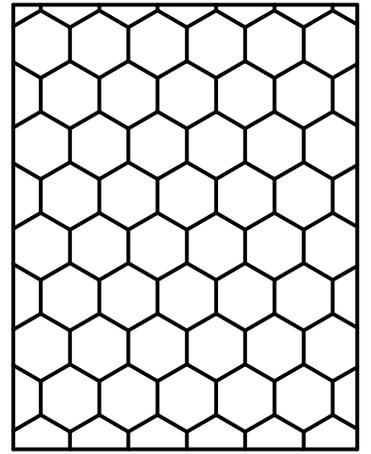
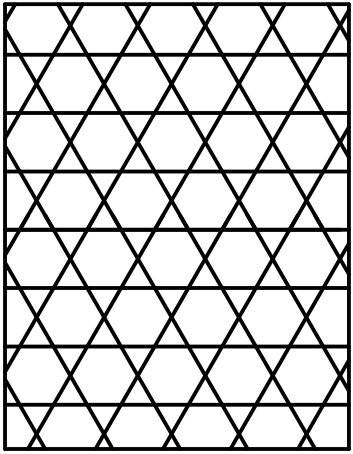
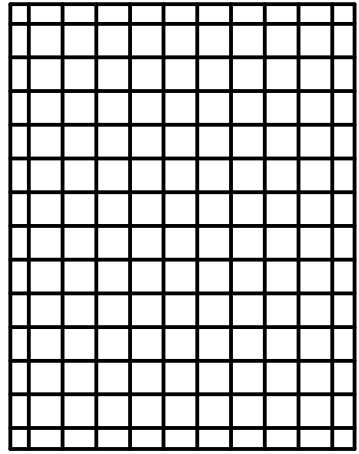
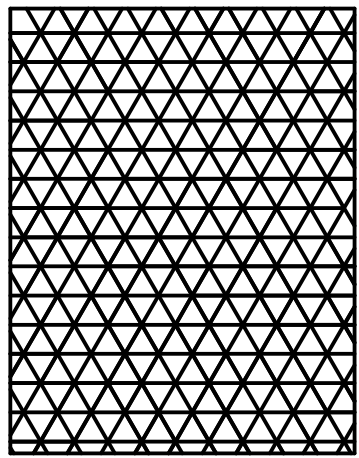
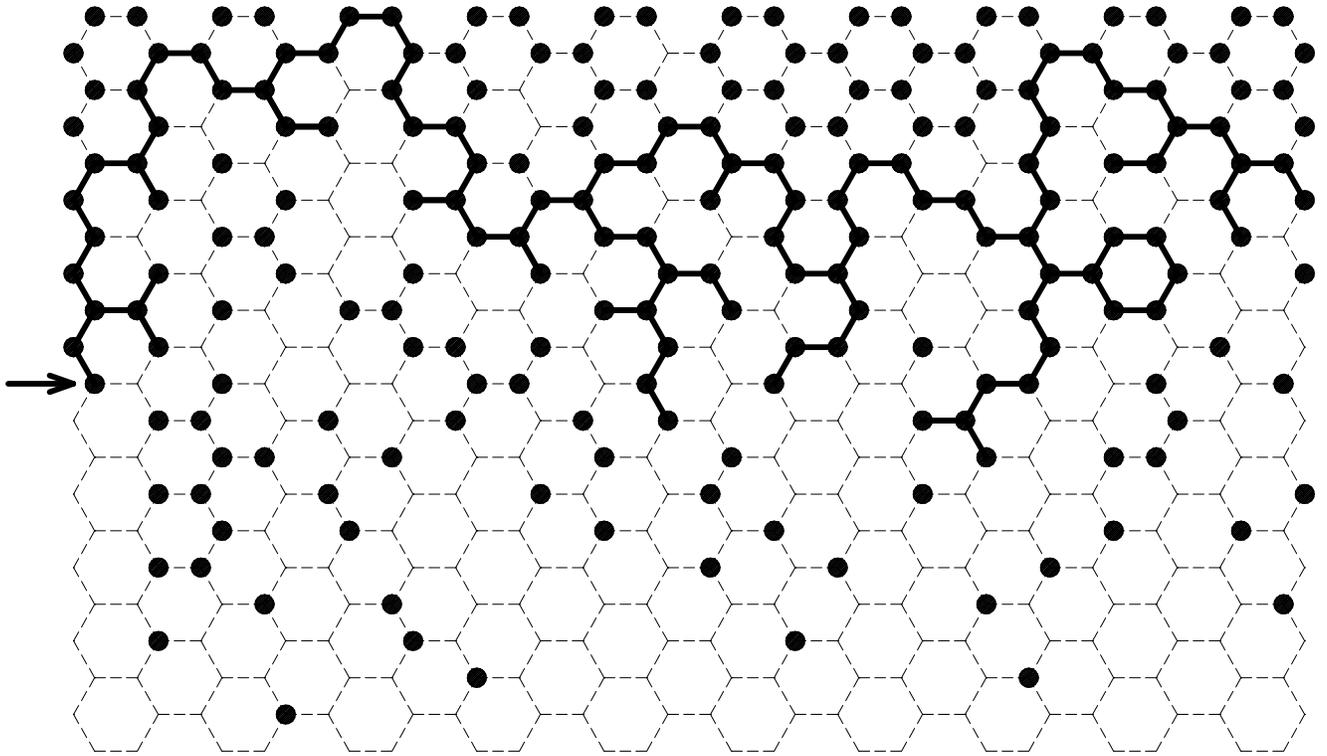
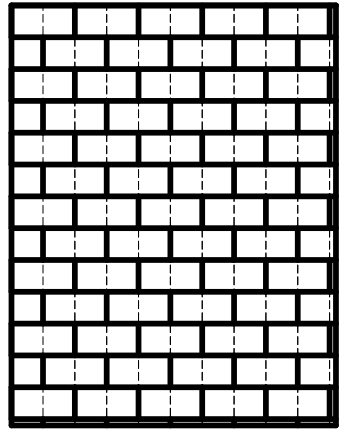
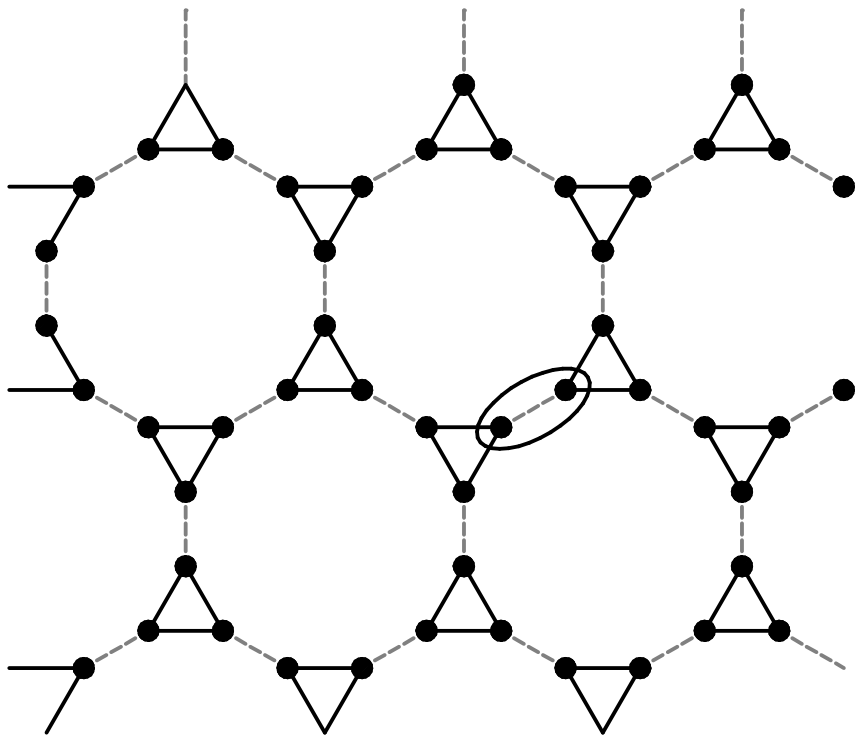
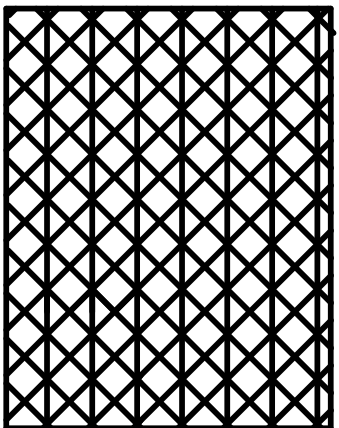
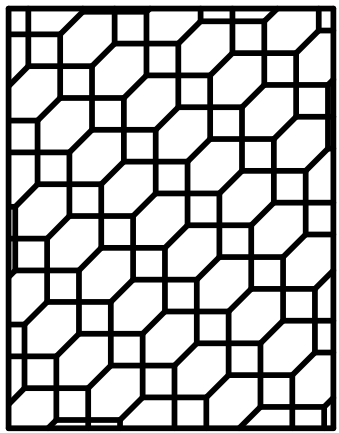
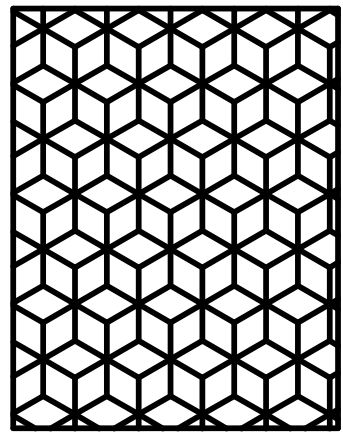
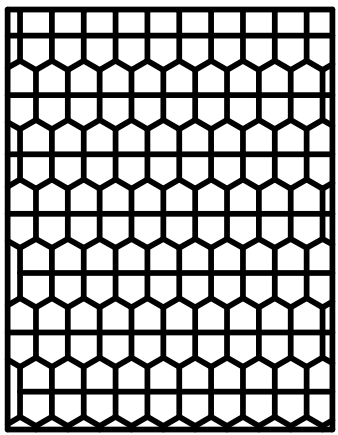
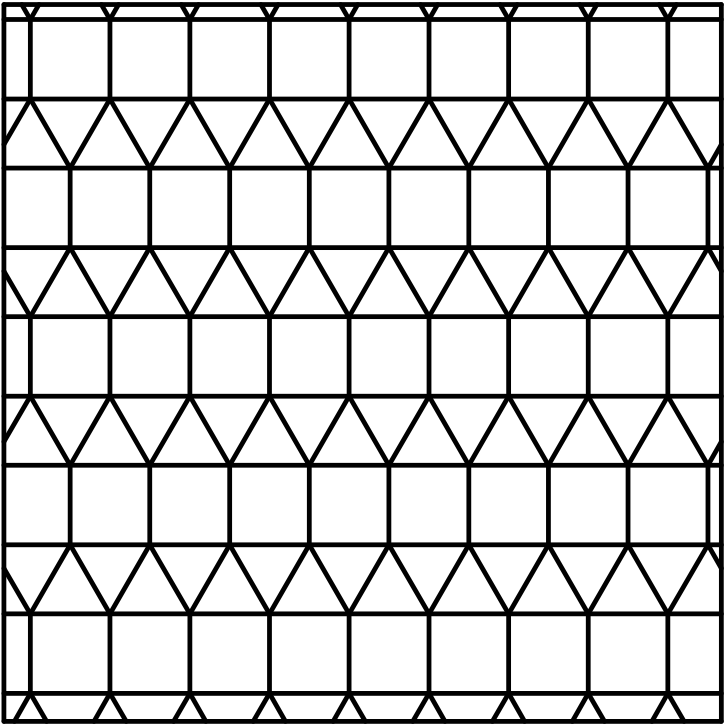
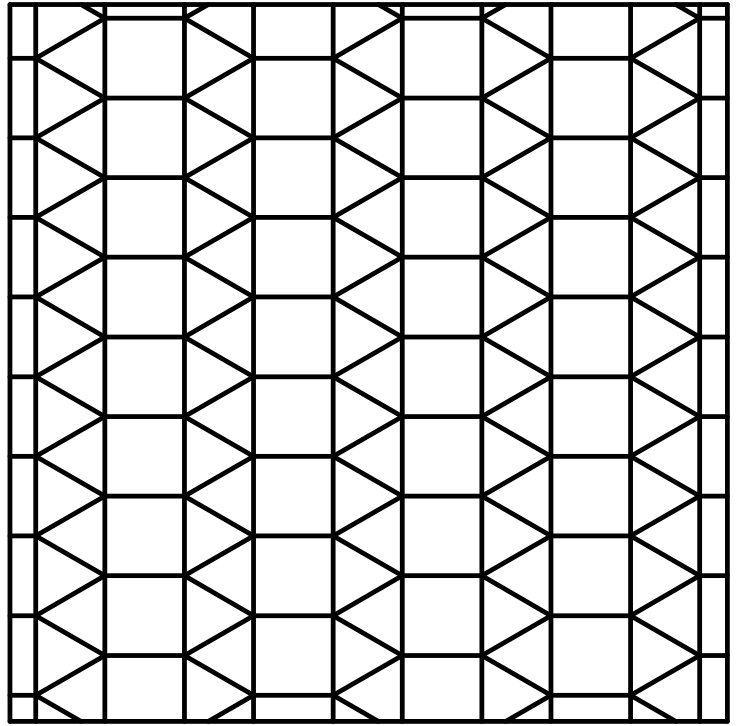
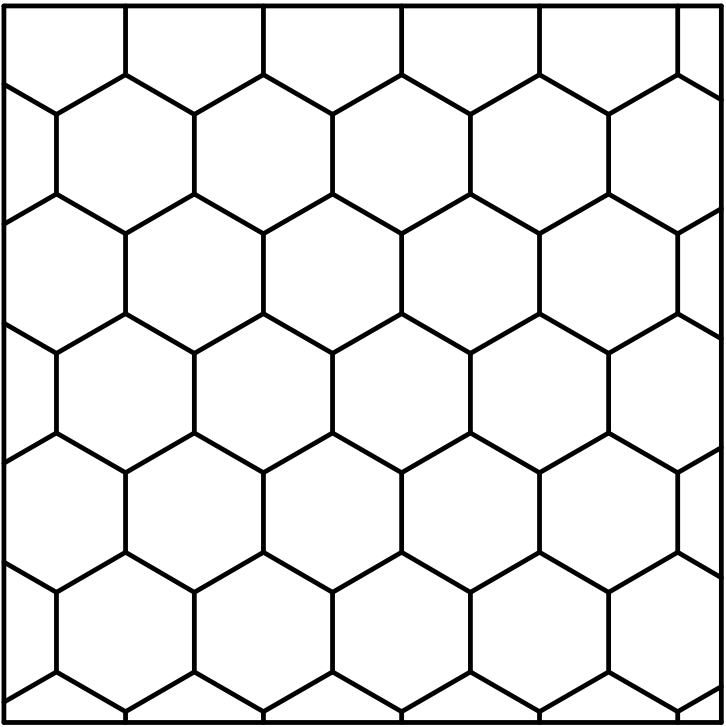
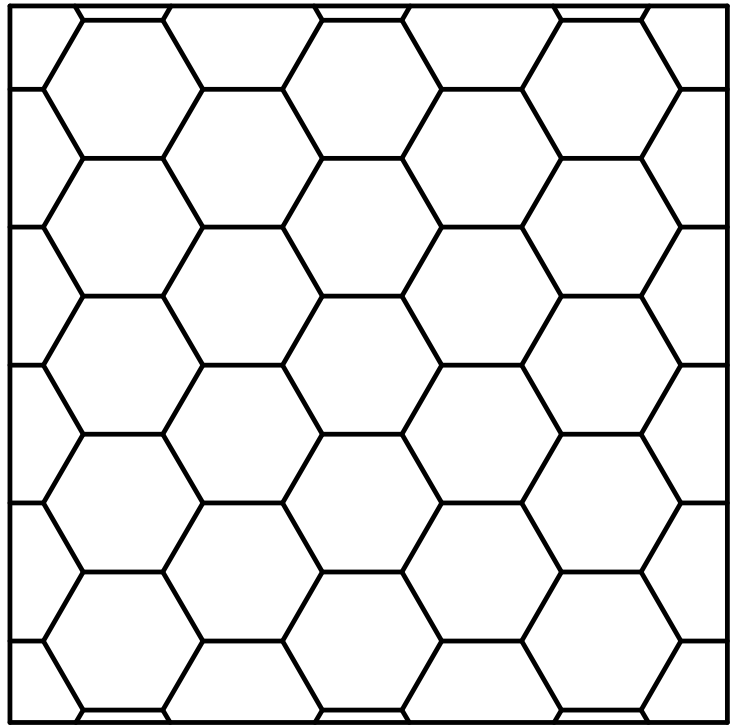
\end{document}